\documentclass{appolb}
\usepackage{epsfig}

\begin{document}
\title{Vector meson production in the dimuon channel in the ALICE experiment at the LHC %
\thanks{\scriptsize{Presented at Strangeness in Quark Matter 2011, 18-24 Sept. 2011, Cracow, Poland}}%
}
\author{L. Massacrier \footnote{\scriptsize{Now at Subatech (Universit\'e de Nantes, Ecole des Mines and CNRS/IN2P3), Nantes, France}} for the ALICE Collaboration
\address{Universit\'e de Lyon, Universit\'e Lyon 1, CNRS/IN2P3, Institut de Physique Nucl\'eaire de Lyon, Villeurbanne}
}
\maketitle
\begin{abstract}
The purpose of the ALICE experiment at the LHC is the study of the Quark Gluon Plasma (QGP) formed in ultra-relativistic heavy-ion collisions, a state of matter in which quarks and gluons are deconfined. The properties of this state of strongly-interacting matter can be accessed through the study of light vector mesons ($\rho$, $\omega$ and $\phi$). Indeed, the strange quark content ($s\bar{s}$) of the $\phi$ meson makes its study interesting in connection with the strangeness enhancement observed in heavy-ion collisions. Moreover, $\rho$ and $\omega$ spectral function studies give information on chiral symmetry restoration. Vector meson production in pp collisions is important as a baseline for heavy-ion studies and for constraining hadronic models. We present results on light vector meson production obtained with the muon spectrometer of the ALICE experiment in pp collisions at $\sqrt{s}$~=~7 TeV. Production ratios, integrated and differential cross sections for $\phi$ and $\omega$ are presented. Those results are extracted for $p_{\rm T} > 1$ GeV/$c$ and $2.5 < y < 4$.
\end{abstract}


Light vector mesons ($\rho$, $\omega$, $\phi$) are interesting in order to study non-perturbative Quantum Chromodynamics (QCD) processes in pp collisions at LHC. In this unexplored energy domain, models of particle production need to be tuned. Moreover, pp collisions are also used as a baseline for heavy-ion collisions where light vector mesons can provide information on the hot and dense state of strongly interacting matter. Study of the $\phi$ meson with respect to the production of $\rho$ and $\omega$ allows for the measurement of strangeness enhancement in heavy ion collisions \cite{RAF,TOUN}. Thanks to its short lifetime, the $\rho$ meson decays inside the QGP, and any modification of the $\rho$ meson spectral function due to chiral symmetry restoration can be then measured \cite{Rapp}.
The results reported here concern data collected for pp collisions at $\sqrt{s}$ = 7 TeV. The analysis of pp and Pb-Pb collisions at \linebreak[4] $\sqrt{s} = 2.76$ TeV is ongoing.\newline


Three detectors of the ALICE experiment \cite{ALICE} were used to perform this analysis. 
The muon forward spectrometer covers the pseudo-rapidity range \linebreak[4] $-4~<~\eta~<~-2.5$~. It consists of a $10\lambda_{I}$ (55 $X_{0}$) thick passive front hadron absorber, five tracking stations composed of two cathod pad chamber (CPC) planes each, a dipole magnet delivering an integrated field of 3~T$\cdot$m, an iron wall that filters secondary and punch-through hadrons, and two triggering stations composed of two resistive plate chamber (RPC) planes each. The spatial resolution of the CPC is $\sim$ 100 $\mu$m in the bending direction while the time resolution of the trigger chambers is 2 ns. Because of the front absorber and the iron wall, muons that reach the trigger stations have a momentum greater than $\sim$ 4 GeV/c.

The Silicon Pixel Detector (SPD) which is the inner part of the silicon vertex tracker of ALICE consists of two layers covering $\mid \eta \mid <$ 2.0 and $\mid \eta \mid <$ 1.6, respectively. In this analysis, the SPD is used for minimum bias triggering and to determine the vertex position of the pp collision. 

Finally, the VZERO detector, made of two arrays of scintillators located at both sides of the interaction point, covers the pseudo-rapidy regions -3.7~$<~\eta~<$~-1.7 and 2.8 $< \eta <$ 5.1. The VZERO was also used for minimum bias triggering and to reject beam-gas interactions thanks to its good timing resolution (1 ns). The ALICE minimum bias trigger condition is given by a logical OR between the requirement of having at least one fired trigger chip in the SPD and a signal in either one of the two VZERO arrays, in coincidence with the presence of the beam bunches. In addition, the muon trigger requires a minimum bias trigger in coincidence with the detection of one muon in the angular acceptance of the muon spectrometer.\newline

\begin{figure}[!htbp]
\vglue -3.7 true cm
\begin{center}
\begin{minipage}[t]{.47\linewidth}
\hglue -0.5 true cm
 \epsfig{file=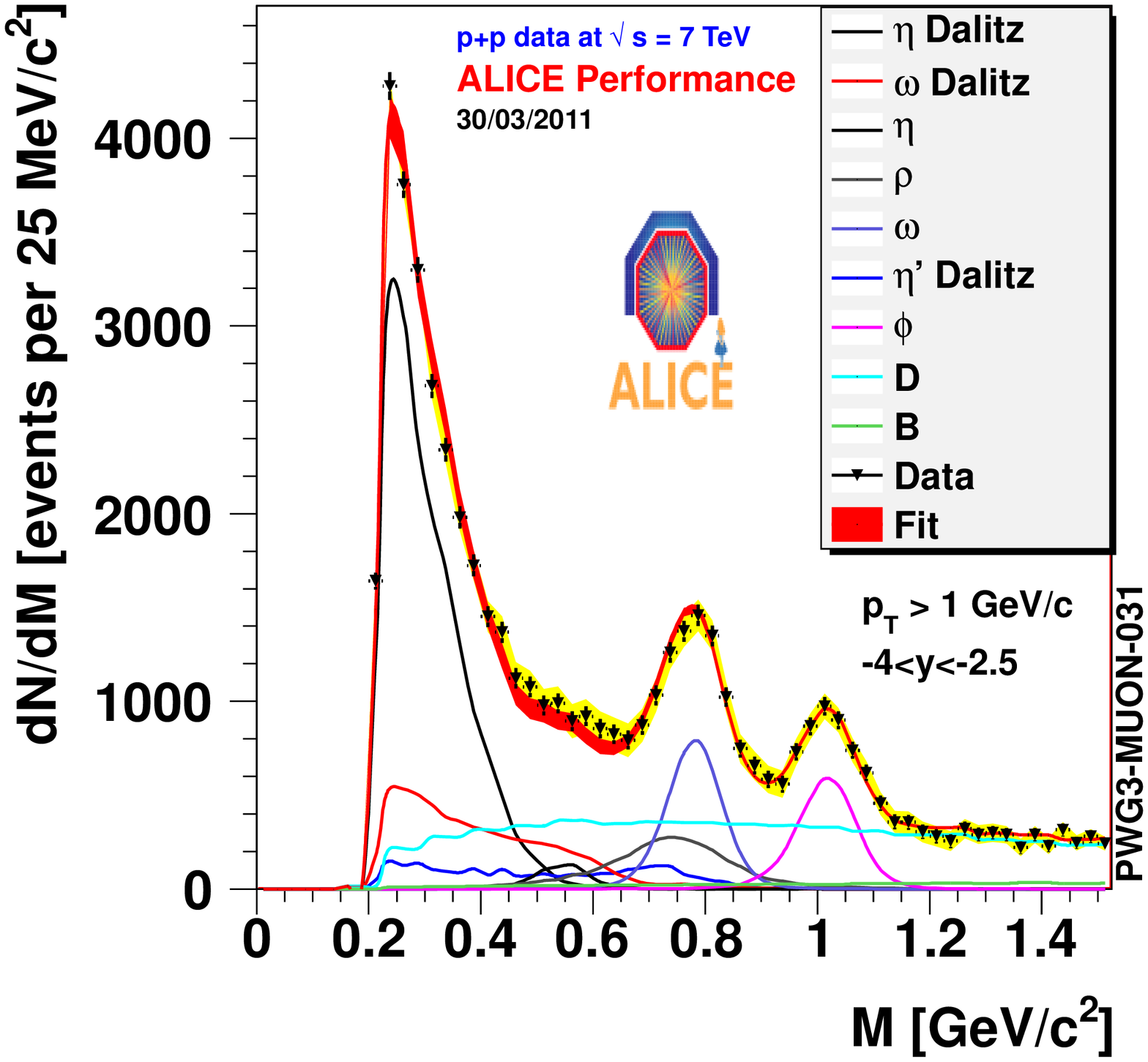, height = 6.5 cm}
\end{minipage}
\begin{minipage}[t]{.47\linewidth}
  \epsfig{file=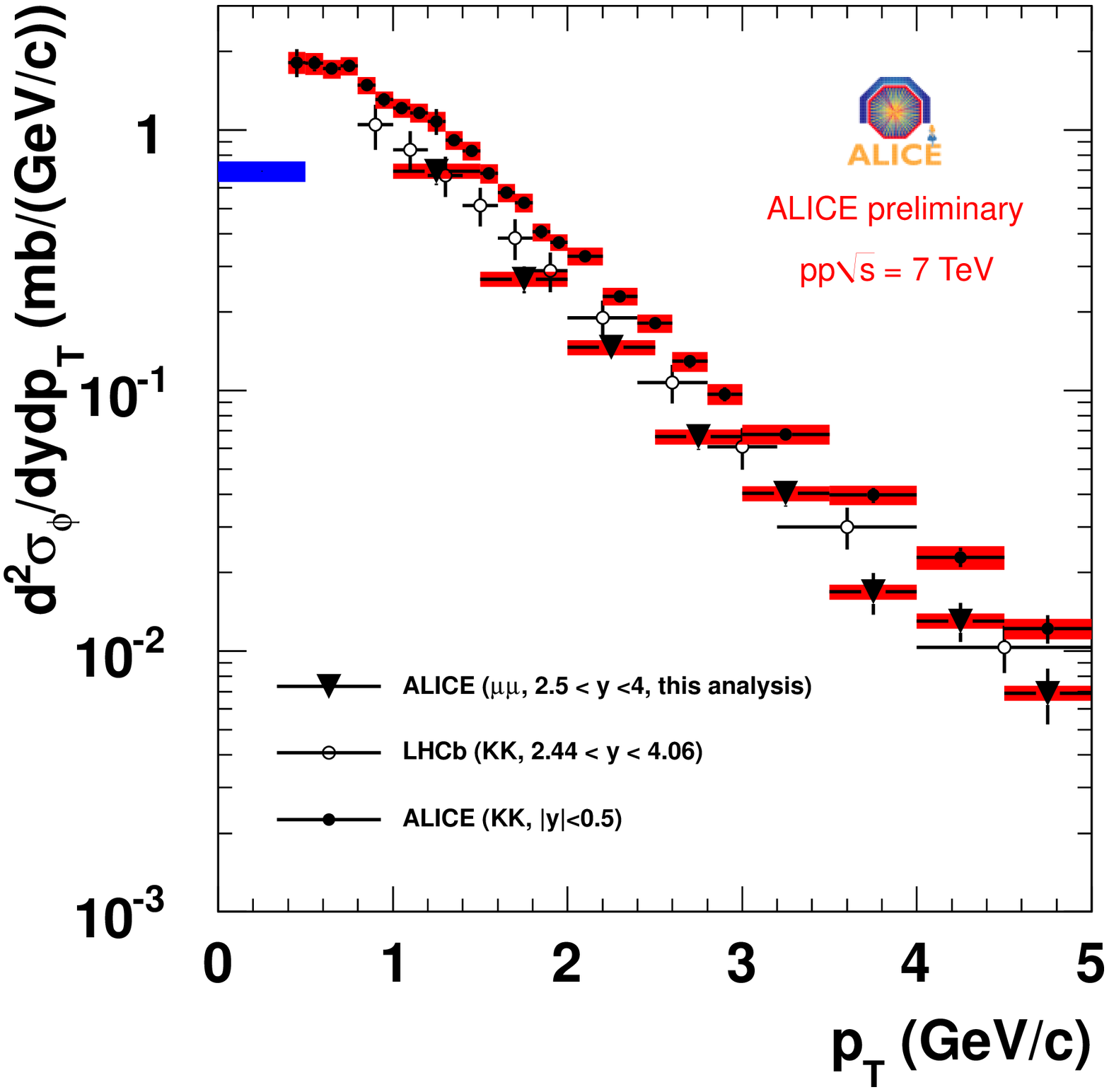, height = 6.5 cm}
\end{minipage}
\caption[res1]{\label{res1} \scriptsize{ Left : Dimuon invariant mass spectrum in pp collisions at $\sqrt{s}$ = 7 TeV after background subtraction (event mixing). Simulations of all the processes involved in the low mass region are performed with a hadronic cocktail generator to fit the data. The yellow band represents the systematic uncertainties on background subtraction. The red band are the errors on the fixed values of the fit. Right : Inclusive $\phi$ differential production cross section $d^{2}\sigma_{\phi}$/$dp_{\rm T}dy$ in pp collisions at $\sqrt{s}$ = 7 TeV and for 2.5$ < y < $4. Black triangles correspond to the measurement performed with the muon spectrometer of ALICE in the dimuon decay channel. The blue box corresponds to the error on normalization. Red boxes are the point to point uncorrelated systematic uncertainties. The data are compared to the measurement in the kaon decay channel done by ALICE in the central barrel (solid circles) \cite{PULV} and by LHCb (black open circles) \cite{LHCb}.}}
\end{center}
\end{figure}

The analysed data sample in pp collisions at $\sqrt{s}$ = 7 TeV corresponds to an integrated luminosity of approximatively 85 nb$^{-1}$ for the extraction of the $p_{\rm T}$ distributions. Part of these data are discarded for cross sections measurements to avoid running periods with a high pileup rate. The remaining data sample corresponds to an integrated luminosity of 56 nb$^{-1}$. 
Each muon track reconstructed in the tracking chambers is required to match a tracklet in the trigger chambers and a cut on the muon rapidity, $2.5 < y_{\mu} < 4$, is applied to remove muons at the edge of the acceptance. \\
Several processes contribute to the dimuon invariant mass spectrum in the low mass region. The sources which generate correlated dimuons are the following: $\eta \rightarrow \mu^{+}\mu^{-}$, $\rho \rightarrow \mu^{+}\mu^{-}$, $\omega \rightarrow \mu^{+}\mu^{-}$, $\phi \rightarrow \mu^{+}\mu^{-}$, the Dalitz decays of $\eta$, $\omega$ and $\eta'$ trough the processes $\eta \rightarrow \mu^{+}\mu^{-}\gamma$, $\eta' \rightarrow \mu^{+}\mu^{-}\gamma$, $\omega \rightarrow \mu^{+}\mu^{-}\pi^{0}$ and the semi-leptonic decays of open charm and open beauty. In addition to those processes, the semi-leptonic decays of pions and kaons produce uncorrelated dimuons. 
This combinatorial background is evaluated using two different techniques. 
The first method is an event mixing technique: single muons from different events are mixed to produce uncorrelated muon pairs. The cuts on those muons are identical to the ones used for invariant mass calculation. The normalization of the spectrum is given by 2$R\sqrt{N_{+ +}N_{- -}}$ where $N_{+ +}$ ($N_{- -}$) is the number of positive (negative) like sign pairs integrated over the whole mass range, while the $R$ factor accounts for any difference in the acceptances of the like and unlike sign muon pairs. It can be evaluated either via simulations, using the relation $R = A_{+ -}/ \sqrt{A_{+ +} A_{- -}}$ where $A_{+ -}$, $A_{+ +}$, $A_{- -}$ are the acceptances of opposite sign, positive like sign and negative like sign muon pairs, or from the mixed events, through the relation $R = N^{mixed}_{+ -}$/$\sqrt{N^{mixed}_{+ +} N^{mixed}_{- -}}$ where $N^{mixed}_{+ -}$, $N^{mixed}_{+ +}, N^{mixed}_{- -}$ are the number of mixed unlike sign and like sign muon pairs. The second method for background estimation uses the real like sign muon pairs. In this second approach, the number of background pairs for a given bin of mass ($\Delta M$) is calculated with the formula: $N^{+ -}(\Delta M) = 2 R (\Delta M) \sqrt{N_{++}(\Delta M) N_{- -} (\Delta M)} $ where $N_{++}(\Delta M)$,  ($N_{- -}(\Delta M)$) is the number of positive (negative) like sign pairs for a given bin of mass. Results from the two methods used to determine the combinatorial background agree for $p_{\rm T}$ of the muon pair above 1 GeV/$c$ in the mass range 0.3 to 1.5 GeV/$c^{2}$. 
The validity of the event mixing is checked by comparing the like sign mixed pairs with the like sign unmixed pairs. If the like sign pairs from data are purely uncorrelated, the mixed pairs shall reproduce the unmixed pairs spectrum. The two spectra show similar trend. Their ratio is equal to unity within 5$\%$. This value is taken as a systematic uncertainty on the background normalization. In the following, the background subtraction is performed with the event mixing technique because of the large statistics of mixed pairs available. The invariant mass spectrum of unlike sign muon pairs after combinatorial background subtraction is shown in Fig. \ref{res1} left.
Simulations of all the correlated dimuon sources are performed with an hadronic cocktail generator in order to fit the invariant mass spectrum. The free parameters of the fit are the normalizations of $\eta$ Dalitz, $\omega$, $\phi$ and open charm. The $\rho$ contribution is fixed with the assumption $\sigma_{\rho}$/$\sigma_{\omega}$ = 1. The contribution of other processes is also fixed.\newline

\begin{figure}[!htbp]
\begin{center}
\begin{minipage}[t]{.47\linewidth}
\hglue -0.5 true cm
 \epsfig{file=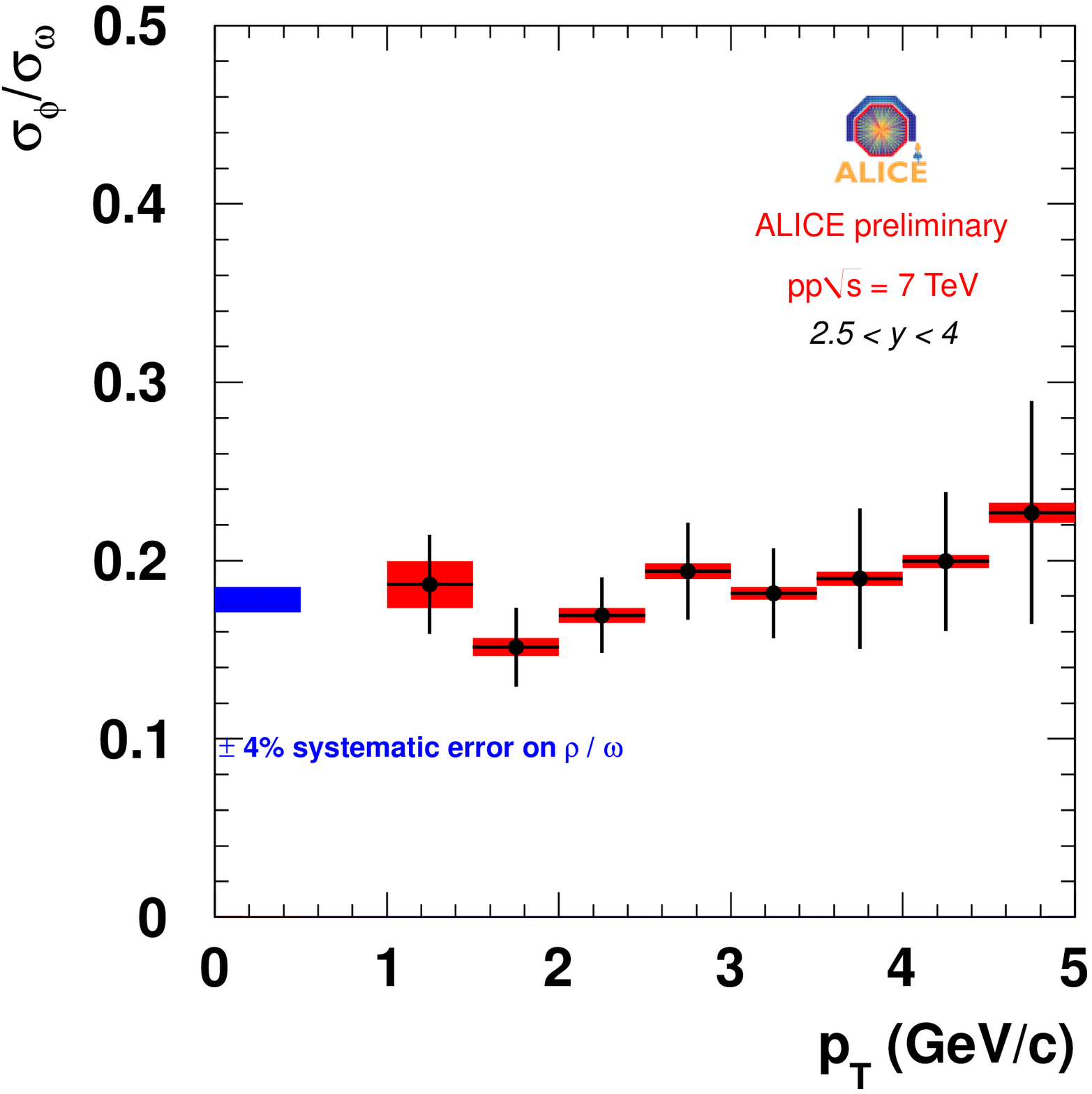, height = 6.5 cm}
\end{minipage}
\begin{minipage}[t]{.47\linewidth}
  \epsfig{file=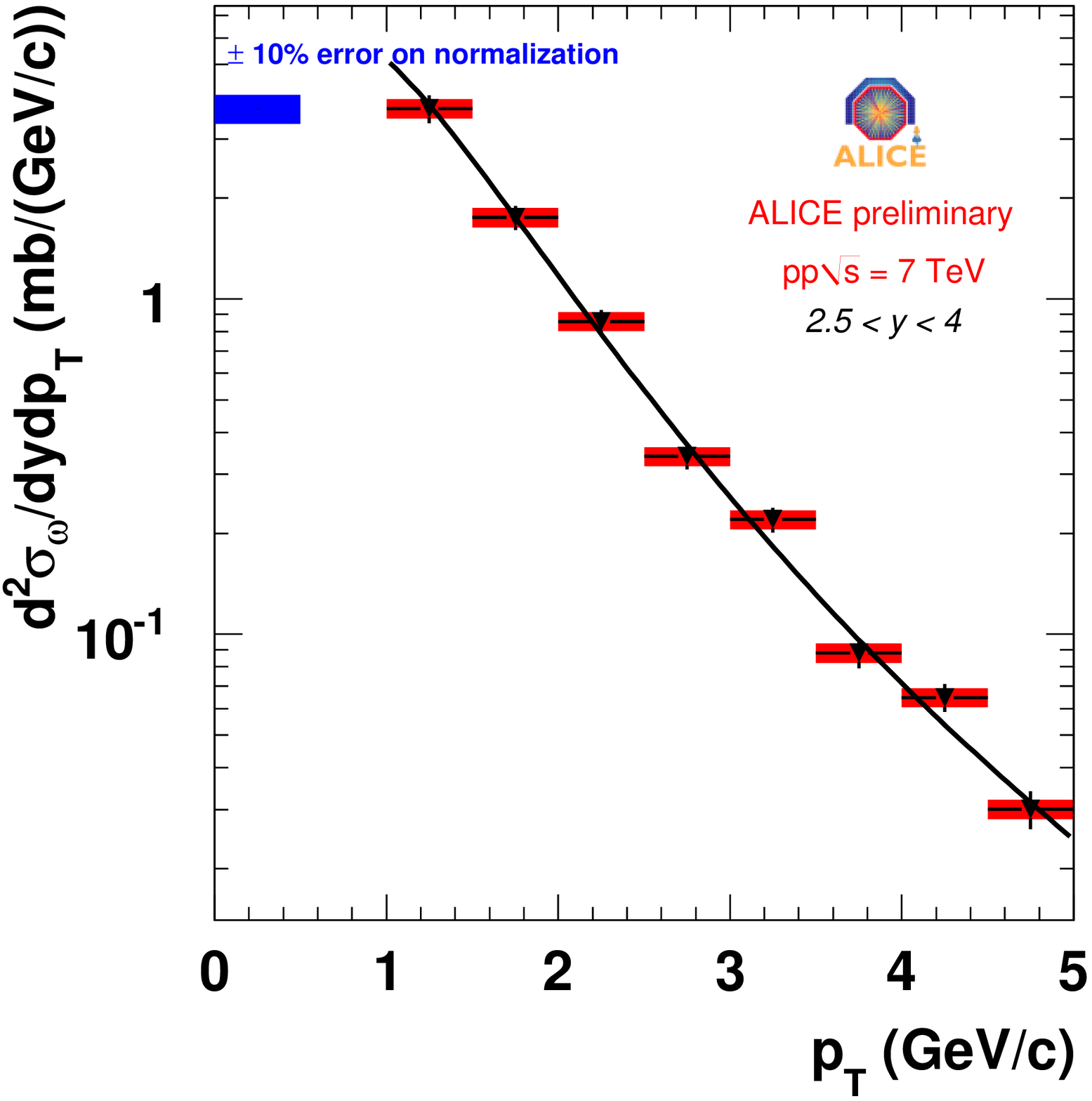, height = 6.5 cm}
\end{minipage}
\caption[res2]{\label{res2}\scriptsize{Left : Ratio of $\phi$ production cross section over $\omega$ production cross section as a function of the transverse momentum in pp collisions at \linebreak[4] $\sqrt{s}$~=~7 TeV. Red boxes are point to point uncorrelated systematic uncertainties and the blue box is the correlated systematic uncertainty. Right : Inclusive differential $\omega$ production cross section $d^{2}\sigma_{\omega}/dp_{\rm T}dy$ for $2.5 < y < 4$ in pp collisions at $\sqrt{s}$ = 7 TeV. The error bars are the quadratic sum of statistical and systematic uncertainties. Red boxes are point to point uncorrelated systematic uncertainties and the blue box is the error on the normalization. A fit is performed with the formula : $C \times p_{\rm T}/[1 + (p_{\rm T}/p_{0})^{2}]^{n}$.}}
\end{center}
\end{figure}

The ratio of $N_{\phi \rightarrow \mu^{+}\mu^{-}}$/$(N_{\rho \rightarrow \mu^{+}\mu^{-}}+N_{\omega \rightarrow \mu^{+}\mu^{-}})$ corrected for acceptance and efficiency is obtained. The average value found for this ratio is 0.416 $\pm$ 0.032 (stat) $\pm$ 0.004 (syst). The main sources of systematic uncertainties are the normalization of the $\omega$ Dalitz, $\eta$' Dalitz and combinatorial background. The $\phi$ cross section is evaluated in the ranges $1~<~p_{\rm T}~<~5$~\hspace{0.1cm}GeV/$c$ and \linebreak[4] $2.5 < y < 4$ with the formula $\sigma_{\phi}~=~\frac{N_{\phi}^{c} \sigma_{MB} N_{\mu}^{MB}}{BR(\phi\rightarrow l^{+}l^{-}) N_{MB} N_{\mu}^{\mu-MB}}$ where $N^{c}_{\phi}$ is the number of $\phi$ corrected for acceptance and efficiency and BR($\phi~\rightarrow ~l^{+}l^{-}$) is the weighted average between dimuon and dielectron branching ratios. $\sigma_{MB}$ is the ALICE minimum bias cross section in pp collisions at \linebreak[4] $\sqrt{s}$ = 7 TeV. In order to get $\sigma_{MB}$ value, the cross section $\sigma_{V0AND}$ is obtained through the measurement of the occurence of a signal in coincidence in the two arrays of VZERO \cite{GAGL} with a van der Meer scan \cite{MEER}. The fraction of minimum bias events where the L0 trigger decision is given by the V0AND is proportional to the ratio $\sigma_{V0AND}$/$\sigma_{MB}$ and was measured to be 0.87 with a stability of 1$\%$ over the whole data taking period. This leads to the measurement $\sigma_{MB} = 62.3 \pm 0.4 (stat) \pm 4.3 (syst)$ mb \cite{OYA}. 
 $N_{MB}$ is the number of minimum bias events corrected for pileup pp collisions in the same bunch crossing. $N_{\mu}^{MB}/N_{\mu}^{\mu-MB}$ is the ratio between the number of single muons collected with the minimum bias trigger and the number of single muons collected with the muon trigger. We obtain $\sigma_{\phi}(1 < p_{\rm T} < 5$ GeV/$c$, $2.5 < y < 4 )$ = 0.940 $\pm$ 0.084 (stat) $\pm$ 0.095 (syst) mb. The different contributions to the systematic uncertainties on the $\phi$ cross section amount to 2$\%$ for the background subtraction, 4$\%$ and 3$\%$ for the trigger and tracking efficiency, respectively, 1$\%$ for the uncertainty on branching ratios, 3$\%$ for the ratio $N_{\mu}^{MB}/N_{\mu}^{\mu-MB}$. The main source of correlated systematic errors comes from the minimum bias cross section (7$\%$). The $p_{\rm T}-$differential cross section $d^{2}\sigma_{\phi}/dp_{\rm T}dy$ is measured and fitted by the power law $C \times p_{\rm T}/[1 + (p_{\rm T}/p_{0})^{2}]^{n}$ which leads to $p_{0}$ = 1.16 $\pm$ 0.23 GeV/$c$ and n= 2.7 $\pm$ 0.2. This $p_{\rm T}-$differential cross section is represented by triangles in Fig. \ref{res1} right. Point to point uncorrelated systematic uncertainties are represented as red boxes. The fully correlated uncertainty is represented by the blue box and it amounts to 9$\%$. Data are compared to ALICE preliminary results at central rapidity (black circles) \cite{PULV} and LHCb measurement (empty circles)\cite{LHCb} in the channel $\phi\rightarrow$ KK. The ALICE central barrel measurement is performed at midrapidity ($\mid$ y $\mid <$ 0.5) while LHCb one in the rapidity region  $2.44 < y < 4.06$, close to our measurement. The three measurements present similar slopes. The differential cross section of $\phi$ meson production measured by LHCb is integrated over $p_{\rm T} >$ 1 GeV/$c$ and extrapolated to the rapidity range covered by our measurement in the dimuon decay channel. The extrapolated value is $\sigma_{\phi}(1 <p_{\rm T} < 5$ GeV/$c$, $2.5 < y <4)$ = 1.07 $\pm$ 0.15~(full error)~mb. The difference between the LHCb and the ALICE dimuon measurement amounts to 14$\%$. The two results are compatible when considering statistical uncertainties and uncorrelated systematics between the two experiments.\\
\indent{The extraction of the $\omega$ cross section requires to disentangle the contributions from the $\rho$ and the $\omega$. Thus the normalization of the $\rho$ is let free in the fit of the mass spectrum. This leads to a measurement for 1 $< p_{\rm T} <$ 5 GeV/$c$ of $\sigma_{\rho}$/$\sigma_{\omega}$ = 1.15 $\pm$ 0.20 (stat) $\pm$ 0.12 (syst) and it is compatible with the hypothesis previously used in the fit. The systematic uncertainties are calculated by modifying the normalization of $\eta$' and $\omega$ Dalitz, taking into account the uncertainties on the branching ratios and by changing the background normalization by $\pm$ 10$\%$. From this measurement and the measurement of $N_{\phi \rightarrow \mu^{+}\mu^{-}}$/$(N_{\rho \rightarrow \mu^{+}\mu^{-}}+N_{\omega \rightarrow \mu^{+}\mu^{-}})$ one can extract the ratio of $\sigma_{\phi}$/$\sigma_{\omega}$ as a function of $p_{\rm T}$ (Fig. \ref{res2} left) with an average value of $\sigma_{\phi}$/$\sigma_{\omega}$ = 0.178 $\pm$ 0.015 (stat) $\pm$ 0.008 (syst). The main source of correlated uncertainties comes from the errors on the ratio $\sigma_{\rho}$/$\sigma_{\omega}$. The extracted $\omega$ production cross section is $\sigma_{\omega}(1 < p_{\rm T} < 5$ GeV/$c$, $2.5 < y < 4$) = 5.28 $\pm$ 0.46 (stat) $\pm$ 0.58 (syst) mb.
In Fig. \ref{res2} right, we present the $\omega$ differential cross section, fitted by the same power law used for the $\phi$ differential cross section. The parameters are found to be $p_{0}$ = 1.44 $\pm$ 0.09 GeV/$c$ and n = 3.2 $\pm$ 0.1.\newline}



We have presented the results on light vector mesons  obtained at \linebreak[4] $\sqrt{s}$ = 7 TeV in pp collisions. The ratio of $N_{\phi}/(N_{\rho}+N_{\omega})$ was measured for $1 < p_{\rm T} < 5$ GeV/$c$ and the average value extracted. The $\phi$ production cross section was extracted for $1 < p_{T} < 5$ GeV/$c$, $2.5 < y < 4$ and is in agreement with the value obtained by LHCb extrapolated to the same rapidity and $p_{\rm T}$ range. The differential cross section of the $\phi$ was obtained and presents a similar slope as the one measured in the ALICE central barrel and LHCb. The $\omega$ cross section and differential cross section were also measured. A similar analysis is ongoing in pp collisions at \linebreak[4] $\sqrt{s}$ = 2.76 TeV to extract the $\phi$ and $\omega$ cross sections.

\end{document}